\documentclass[10pt,letterpaper]{article}
\usepackage[top=0.85in,left=2.75in,footskip=0.75in]{geometry}

\usepackage{amsmath,amssymb}

\usepackage{changepage}

\usepackage[utf8x]{inputenc}

\usepackage{textcomp,marvosym}

\usepackage{cite}

\usepackage{nameref,hyperref}

\usepackage{microtype}
\DisableLigatures[f]{encoding = *, family = * }

\usepackage[table]{xcolor}

\usepackage{color}
\definecolor{orange}{RGB}{255,90,20}
\definecolor{blue}{RGB}{20,90,255}

\usepackage{array}

\newcolumntype{+}{!{\vrule width 2pt}}

\newlength\savedwidth

\raggedright
\usepackage[aboveskip=1pt,labelfont=bf,labelsep=period,justification=raggedright,singlelinecheck=off]{caption}

\makeatletter
\renewcommand{\@biblabel}[1]{\quad#1.}
\makeatother

\usepackage{lastpage,fancyhdr,graphicx}
\usepackage{epstopdf}
\fancyheadoffset[L]{2.25in}
\fancyfootoffset[L]{2.25in}
\begin{document}
\vspace*{0.2in}

\begin{flushleft}
{\Large
\textbf\newline{Predicting risk of dyslexia with an online gamified test} 
}

\vspace*{0.2in}
Luz Rello,\textsuperscript{1*} 
Ricardo Baeza-Yates,\textsuperscript{2} 
Abdullah Ali,\textsuperscript{3} 
Jeffrey P. Bigham,\textsuperscript{4} 
Miquel Serra\textsuperscript{5}\\ 
\bigskip
\textsuperscript{1} Department of Information Systems and Technology, IE University, Madrid, Spain \\

\textsuperscript{2} KCCIS, Northeastern University at Silicon Valley, San Jose, CA, USA 
\\
\textsuperscript{3} School of Computer Science, University of Washington, Seattle, WA, USA
\\
\textsuperscript{4} Human-Computer Interaction Institute, Carnegie Mellon University, Pittsburgh, PA, USA \\
\textsuperscript{5} Department of Cognition and Development, University of Barcelona, Barcelona, Spain
\\
\bigskip

\textsuperscript{1} Roles: Conceptualization, Methodology, Investigation, Data Curation, Writing - Original Draft Preparation.
\textsuperscript{2} Roles: Formal Analysis, Methodology Validation, Writing - Review \& Editing.
\textsuperscript{3} Roles: Software.
\textsuperscript{4} Roles: Supervision, Funding Acquisition, Resources, Writing - Original Draft Preparation.
\textsuperscript{5} Roles: Supervision, Writing - Review \& Editing.

* Corresponding author: luzrello@acm.org. This work was done while the first author was working at the Human-Computer Interaction Institute, Carnegie Mellon University, Pittsburgh, USA.

\end{flushleft}

\pagebreak

\section*{Abstract}
Dyslexia is a specific learning disorder related to school failure. Detection is both crucial and challenging, especially in languages with transparent orthographies, such as Spanish. To make detecting dyslexia easier, we designed an online gamified test and a predictive machine learning model. In a study with more than 3,600 participants, our model correctly detected over 80\% of the participants with dyslexia. 
To check the robustness of the method we tested our method using a new data set with over 1,300 participants with age customized tests in a different environment -a tablet instead of a desktop computer- reaching a recall of over 72\% for the class with dyslexia for children 9 years old or older. 
Our work shows that dyslexia can be screened using a machine learning approach. An online screening tool based on our methods has already been used by more than 200,000 people.

\section*{Introduction}
More than 10\% of the world population has a specific learning disability with neurobiological origin called dyslexia. According to the International Dyslexia Association, ``dyslexia is characterized by difficulties with accurate and/or fluent word recognition and by poor spelling and decoding abilities. These difficulties typically result from a deficit in the phonological component of language that is often unexpected in relation to other cognitive abilities and the provision of effective classroom instruction'' \cite{Lyon2003}. If someone knows they have dyslexia, they can learn coping strategies to overcome its negative effects \cite{krafnick2011gray,Gabrieli2009}. However, when people with dyslexia are not provided with appropriate support, they often fail in school: 35\% of drop out of school, and it is estimated that less than 2\% of people with dyslexia will complete a four year college degree \cite{Al-Lamki2012}.


Detecting dyslexia is especially difficult in languages with transparent orthographies, such as Spanish. In such languages, the correspondence of grapheme (letter) and phoneme (sound) is more consistent than in languages with deep orthographies, such as English, where people with dyslexia struggle more in learning how to read \cite{Vellutino2004,Brunswick2010} and thus dyslexia is easier to detect.
Therefore, dyslexia is called a ``hidden disability'' due to the difficulty of its diagnosis in languages with transparent orthographies because manifestations of dyslexia are less severe \cite{Vellutino2004,Brunswick2010}. As a result, Spanish speakers primarily learn that they might have dyslexia through school failure,
which is often too late for effective intervention. Current methods of diagnosis and screening require professionals to collect performance measures related to reading and writing via a lengthy in-person test \cite{PROLECR,TALE1984,DSTJ}, measuring, {\em e.g.}, reading speed (words per minute), reading errors, writing errors, reading words, pseudo-word reading, reading fluency or text comprehension.

While machine learning techniques are broadly used in medical diagnosis  \cite{kononenko2001machine}, in the case of dyslexia it has been only used in combination with eye-tracking measures \cite{W4A2015,benfatto2016screening}. 
The scope of this study is to determine whether people with and without dyslexia can be screened by using machine learning with input data from the interaction measures when being exposed to gamified linguistic questions through an online test, so it is easier to administer.

\section*{Materials and Method}
\subsection*{Method} 
We designed 32 linguistic exercises appropriate for inclusion into a web-based gamified test and conducted a study with 3,644 participants (392 with professional dyslexia diagnosis). Using a within-subject experimental design, we collected numerous performance measures during test completion.

\subsection*{Content Design} 
We designed the gamified exercises using two methods. First, some exercises were based on an empirical analyses of a corpus or errors written in Spanish by people with dyslexia  \cite{LRE2016} because errors reflect specific difficulties that comprise dyslexia \cite{afonso2015spelling,suarez2014orthographic}: we annotated the mistakes with general linguistic characteristics as well as with phonetic and visual information \cite{LRE2016}. We then analyzed the mistakes and extracted statistical patterns to later use in the creation of the test questions. Examples of these patterns are found in the most frequent linguistic and visual features shared in the errors which are phonetically and visually motivated.
For instance, the most frequent errors involve letters in which the one-to-one correspondence between graphemes (letters) and phonemes (sounds) is not maintained, such as ($<$b, v$>$, $<$g, j$>$, $<$c, z$>$, $<$c, s$>$, $<$r$>$) and the letter $<$h$>$, which, in most cases, does not have a phonetic realization in Spanish. Another example of this phonological motivation found in errors is that mistakes involving vowel substitutions take place between phonemes that share one or two phonetic features, with lip rounding being the most frequently involved in errors ($<$a, e$>$). On the other hand, the visual motivation is demonstrated in that 46.91\% of the error letters occur with mirror letters, {\it i.e.}, $<$p$>$ and $<$q$>$ or $<$n$>$ and $<$u$>$ \cite{LRE2016}.

Second, we designed test exercises to target specific cognitive processes, different types of knowledge, and difficulties entailed in reading \cite{Suarez2012reading,Davies2013lexical,suarez2015reading}. Each exercise addresses three or more of the following dyslexia-related indicators shown in Table \ref{indicators} that are different types of {\it Language Skills, Working Memory, Executive Functions} and {\it Perceptual Processes}. 

\begin{table}[ht!]
\begin{center}
\begin{tabular}{|l|l|}
\hline
 {\bf Language Skills}      & {\bf Working Memory} \\ \hline
 Alphabetic Awareness       & Visual (alphabetical) \\
  Phonological Awareness    & Auditory (phonology) \\
 Syllabic Awareness         & Sequential (auditory) \\
  Lexical Awareness         & Sequential (visual) \\
 Morphological Awareness    & {\bf Executive Functions} \\ \cline{2-2} 
  Syntactic Awareness       & Activation and Attention \\
  Semantic Awareness         & Sustained Attention \\ 
  Orthographic Awareness    & Simultaneous Attention    \\ 
   \hline
   \multicolumn{2}{|l|}{{\bf Perceptual Processes}} \\   
   \multicolumn{2}{|l|}{{Visual Discrimination and Categorization}} \\  
   \multicolumn{2}{|l|}{{Auditory Discrimination and Categorization}} \\
   \hline
\end{tabular}
\end{center}
\caption{Cognitive indicators used in the creation of test exercises.}
\label{indicators}
\end{table}

\begin{figure}[ht!]
	\centering
	\includegraphics[width=\columnwidth]{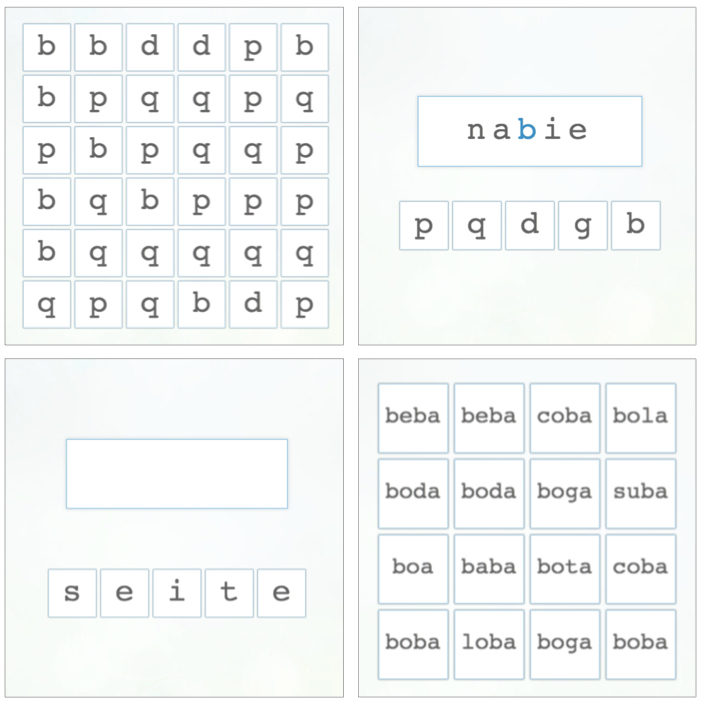}
	\caption{Examples of four test questions: find  `d' among `b', `p', and `q (top left); build a correct word (`nadie', {\it `nobody'}) by substituting one letter (top right);  re-order the letter to write a correct (`siete', {\it `seven'}) (bottom left); and find the word `boda' ({\it `wedding'}) (bottom right). The instructions of the game were given via prerecorded voice prompts.}
	\label{exercises}
\end{figure}

The language of the exercises was reviewed by five speech therapists from Spain, Chile and Argentina, to guarantee that the Spanish variant presented in the exercise was neutral. To ensure that the pronunciation was performed correctly, the voices in the exercises were recorded by a professional voice actress. Likewise, to ensure that the difficulty level was appropriate, each question was reviewed by the speech therapists. See Figure \ref{exercises} for an example of the exercises layout. 

Questions 1-21 (Q1-Q21) entangled auditory and visual discrimination and categorization of different linguistics elements (phonemes -sounds-, graphemes -letters-, syllables, words, pseudo-words). As the level increases, the elements are harder distinguish, because they are phonetically and orthographically more similar. The questions of the test were presented in increasing order of complexity and were intended for children seven years old or older. 

Previous work \cite{afonso2015spelling,suarez2015reading,Davies2013lexical} has shown that people with dyslexia have difficulty recognizing their own reading and spelling errors, including insertion, deletion, substitution or transposition of letters and syllables as well as detecting syntactic and semantic errors in sentences, that is, errors in the structure or in the sentence meaning. Hence, exercises 22-29 (Q22-Q29) focus on correcting words and sentences by fixing the type of errors found in texts written by people with dyslexia. For instance, the user is asked to re-order the letters in the common error *`seite' to form the correct word `siete'{\it `seven'} (See Figure \ref{exercises}). These exercises target lexical knowledge, word identification, reading comprehension, and other linguistics skills such as phonological, syntactic and semantic awareness. 

The final exercises (Q30-Q32) target sequential visual and auditory working memory by asking the player to write sequences of letters in an specific order as well as words and pseudo-words. 

A more detailed explanation of each question can be found in the Data sets section.

The test was implemented in HTML5, CSS and Javascript with a back-end PHP server and a MySQL database.

\subsection*{Participants} 
Children and adults with dyslexia were recruited through a public call to dyslexia centers and dyslexia associations; the inclusion criteria specified that participants should have a dyslexia diagnosis performed by a registered professional. Participants without dyslexia were recruited through schools and limited to children and adults who had never had language problems in school records. Determining accurate ground truth in dyslexia diagnosis is difficult precisely because many people are never diagnosed and we do not know the ground truth accuracy of the professional diagnoses.
All the participants' native language was Spanish.

The participants with dyslexia consisted of 392 people (45.2\% female, 54.8\% male). Their ages ranged from 7 to 17 (M = 10.90, SD = 2.58). The group of participants without dyslexia was composed of 3,252 people (49.7\% female, 50.3\% male), ages ranging from 7 to 17 (M = 10.44, SD = 2.46).

\subsection*{Dependent measures}
To quantify task performance, we collected the following dependent measures for each question: (i) number of {\it Clicks}; (ii) number of correct answers {\it (Hits)} ; (iii) number of incorrect answers {\it (Misses)}; (iv) {\it Score} defined as sum of Hits per set of exercises; (v) {\it Accuracy}, defined as the number of Hits divided by the number of Clicks; (vi) {\it Missrate}, defined as the number of Misses divided by the number of Clicks. 
We later used these performance measures together with the demographic data as features of our prediction model's data set, see section Data sets.

\subsection*{Compliance and Ethics Statements} 
Interested organizations responded to our public calls, and, those where we verified that met the participation requirements (age, mother languages, technical requirements and dyslexia diagnosis for the experimental group) were included. 
Overall, 113 organizations from Argentina, Chile, Colombia, Spain, and USA participated in the study: 3 universities, 60 schools including primary and secondary, 22 specialized centers that support people with dyslexia, and 18 non-profit organizations compose of 4 foundations and 14 associations of dyslexia in Hispanic countries. Most organizations included both, dyslexia and non-dyslexia subjects. For each of the organizations there was one or more supervisors who were trained to administer the study protocol. 

\subsection*{Procedure} 

This study was approved by the Carnegie Mellon University Institutional Review Board (IRB). First, participants gave their written on-line consent. In case the participant was under-aged we gathered consents from the participants and their parents. Then, the participant or the supervisor --in case the participant was under-aged-- filled out a demographic questionnaire, including the date of their dyslexia diagnosis (if any).

Next, they were given instructions on how to fulfill the tasks and completed the study: they completed the gamified test for 15 minutes without interruption. Supervisors could not help participants complete the test using a desktop computer. For schools, parental consent was obtained in advance and the study was supervised by the school counselor or the therapist. All participants and supervisors were volunteers.

\subsection*{Data sets}

We had 3,644 participants with an age range of 7 to 17 years old, where 392 (10.7\%) had diagnosed dyslexia.
In Table~\ref{datasets} we show the characteristics of the overall data set as well as the characteristics of age-based subsets.

\begin{table}[ht!]
\begin{center}
\begin{tabular}{|l|c|c|c|c|c|} \hline
Data set & Participants & Ave. Age & Dyslexia & Female & Male \\ \hline\hline
A1 (7-17)  & 3,644 & 10.90 & 10.8\% & 49.2\% & 50.8\%   \\ \hline
A2 (9-17) & 2,733 & 11.45 & 12.0\% & 49.4\% & 50.6\% \\
A3 (7-11) & 2,539 & 9.12 & 10.1\% & 49.3\% & 50.7\% \\
A4 (9-11) & 1,628 & 9.97 & 11.9\% & 49.6\% & 50.4\% \\
A5 (12-17) & 1,105 & 13.62 & 12.2\% & 49.0\% & 51.0\% \\
A6 (7-8)  & ~~911   & 7.60 & 6.9\% & 48.7\% & 51.3\% \\ \hline
\end{tabular}
\caption{Characteristics of the data sets (age range).}
\label{datasets}
\end{center}
\end{table}

The data for each participant consisted of a total of 196 features: 
Features from 1 to 4 correspond to demographic features, while features from 5 to 196 to the performance features, derived from their interaction while playing the 32 questions of the test (6 measures per question presented previously, that is, {\it Clicks, Hits, Misses, Score, Accuracy,} and {\it Missrate}). Following, we describe them in detail.

\begin{itemize}
\item [1] {\bf Gender} of the participant, a binary feature with two values: {\it Female} and {\it Male}.
\item [2] {\bf Native language} of the participant, a binary feature with two values: {\it Yes} if their native language was Spanish and {\it No} if they were bilingual, being one the languages Spanish.
\item [3] {\bf Language subject}. This is a binary feature with two values: {\it Yes} when the participant had fail a language subject at school at least once and {\it No} when the participant had never fail that subject.
\item [4] {\bf Age} of the participant ranging from 7 to 17 years old.

\item [5-28] These features correspond to questions from 1 to 4 (Q1-Q4).  In these tasks the participant hears the name of a letter ({\it e, g, b,} and {\it d}) and maps it with the letter among  distractors (orthographic and phonetically similar letters) within a time frame, using a Whac-A-Mole game interaction. These questions address prerequisites in reading acquisition: {\it Alphabetic Awareness}, {\it Phonological Awareness} and {\it Visual discrimination and categorization}.

\item [29-58] Features targeting {\it Phonological Awareness}, {\it Syllabic Awareness} and {\it Auditory Discrimination and Categorization}. Here the players hear the pronunciation of syllable ({\it ba}, {\it gar}, {\it pla}, {\it bla} or {\it glis} corresponding to Q5, Q6, Q7, Q8 and Q9, respectively) and map it with its spelling, {\it e.g. glis} where the distractors are {\it glir, glin, gris, gril, grin, glim, gles, grel, glil} and {\it glen}.

\item [59-82] Features corresponding to a set of exercises (Q10-Q13) where participants map a pronunciation of word with its spelling ({\it e.g. boda }) discriminating among other phonetically and orthographically similar words and/or pseudo-words {\it e.g. boba, boca, boga, bola, bota, baba, beba, deba, tuba, buba, suba, loba} or {\it coba}. These features aim at {\it Lexical Awareness}, {\it Auditory Working Memory}, and {\it Auditory Discrimination and Categorization}.

\item [83-106] These performance features (Q14-Q17) target mainly {\it Visual Discrimination and Categorization}, and {\it Executive Functions} since the players undertake a visual search task, finding as many as possible different letters with-in a time frame, {\it e.g. E/F, g/q, u/n c/o, b/d, d/p, b/q}, among others.

\item [107-130] Features extracted from a set of exercises (Q18-Q21) where players listen to a pseudo-word and choose its spelling ({\it e.g. pamata }) among ({\it e.g. mapata, matapa, pamada, mapaba, madata, damata, pamama,} and {\it mamata}). These features target {\it Visual Working Memory}, {\it Sequential Auditory Working Memory}, and {\it Auditory Discrimination and Categorization}.

\item [131-142] These target mainly {\it Lexical}, {\it Phonological}, and {\it Orthographic Awareness}; extracted from exercises where participants need to fill the missing letter in a word, {\it i.e. ha\textunderscore e} for {\it hace} (Q22, features 131-136) or delete the extra letter in the word (Q23, features 137-142) {\it i.e. feiria} for {\it feria}.

\item [143-148] These performance features (Q24) mainly target {\it Morphological} and {\it Semantic Awareness}. They are collected from exercises were participants find a morphological error in a sentence -which gives as a result a semantic error-. For instance, in the sentence {\it Voy a la pastelería a *comparar un pastel. (`I go to the bakery to *compare cake')}, where the word {\it comparar (`to compare')} should be {\it comprar (`to by')}.

\item [149-154] Features related (Q25) to {\it Syntactic Awareness}. Similarly to the previous set of exercises, participants need to find and error in a sentence, being this error in a grammatical or functional word, so the Syntactic meaning of the sentence change. {\it e.g. al (`at')} instead of {\it la (`the')} in {\it Est\'a la final de la sala ('There is *the end of the room')}.

\item [155-160] A set of features (Q26) related to {\it Phonological},  {\it Lexical} and, {\it Orthographic Awareness} since they are extracted from exercises where participant ween to find an error in a word, {\it i.e. *egemplo}, and correct it {\it ejemplo (`example')} choosing a letter from a set of distractors {\it j,n,d,} and {\it b}.
    
\item [161-172] Here participants are asked to rearrange these letters to spell a real word (Q27, for features 161-166), {\it e.g.  `s', 'e', 'i', 't' , 'e'} to build {\it siete (`seven')}  {\it Phonological, Lexical} and  {\it Orthographic Awareness} or to rearrange these syllables to spell a real word (Q28, for features: 167-172), {\it e.g.  'ra','do','mo'} to build {\it morado (`purple')} {\it Syllabic, Lexical} and {\it Orthographic Awareness}.

\item [173-178] These features (Q29) address {\it Phonological}, {\it Lexical} and  {\it Orthographic Awareness} derived from exercises where players separate the words to make a meaningful sentence, {\it e.g. Hoycumploveintid\'osa\~nos} to {\it Hoy cumplo veintid\'os a\~nos (`I'm twenty-two today')}.

\item [179-184] This set of features (Q30) target {\it Sequential Visual Working Memory} since they are gathered from exercises were players see for 3 seconds a sequence of letters {\it ($<$i u a$>$, $<$p g d j$>$, $<$v h b z q$>$,} and {\it $<$M D J N P H$>$)} and then write then discriminating the targets from the distractors {\it Visual discrimination and categorization}.

\item [185-196] These features are derived from dictation tasks where participant listen and write four words (Q31) {\it e.g. principio} 
    ({\it Lexical}, {\it Orthographic Awareness} and {\it Auditory Working Memory}) and four pseudo-words (Q32) {\it e.g. danama} {\it Sequential Auditory Working Memory} and 
    {\it Phonological Awareness}.
\end{itemize}

Since all the exercises involve attention all the performance features [5-196] target the {\it Executive Functions} of {\it Activation and Attention}, and {\it Sustained Attention}. In addition some of them (Q24-Q26) also target {\it Simultaneous Attention} when the participant pays a attention from a number of sources of incoming information at the same time, {\it e.g.}, word recognition, distractor discrimination and error recognition.

\section*{Results} 
\subsection*{Predictive model}
For the predictive model, we used Random Forests \cite{RF2001} due to their non-linearity and good level of interpretability, as this technique is based in decision trees with bagging. We used the Weka 3.8.3 implementation with 200 iterations and unlimited height, obtaining models with a few hundred trees. We did not optimize any parameter to avoid over-fitting, although bagging already minimizes the variance with the same goal in mind.
For all cases we used weighted attributes to balance the dyslexia with the non-dyslexia classes as a trivial classifier (everyone does not have dyslexia) would have obtained an accuracy of 89.2\%, since 10.8\% of the participants have dyslexia.

For the evaluation we used a 10-fold cross validation. That is, we divide the data in 10 random groups (6 of size 364 and 4 of size 365) and then we use 9 of them for training and the last one for validation, repeating this 10 times changing the validation set. This is much better than a 10\% single held-out subset, as we average 10 different partitions instead of just one. This even further avoids over-fitting or random luck in a single partition.

As our goal is to have high recall (sensitivity) for the dyslexia class, we choose the Random Forest voting decision threshold such that the weight of the false negatives is similar to the weight of the false positives. This implies that we give between 8 to 9 times more importance to not send a child with dyslexia to the specialist than sending a child without dyslexia to the specialist. This implies that the threshold will be much less than 0.5, which is the default value. We discuss this issue in the next section.

\begin{table}[ht!]
\begin{center}
\begin{tabular}{|l|c|c|c|c|c|} \hline
Data set & Accuracy & Recall & Precision & ROC & Threshold \\ 
(age range) & (\%) & (Dys., \%) & (Dys.,\%)& & \\ \hline\hline
A1 (7-17) & 79.4 &	80.4 &	79.7  & 0.871 & 0.24 \\ \hline
A2 (9-17) & 80.1 & 79.9 & 80.1	&	0.878 &	0.26 \\
A3 (7-11) & 80.8 &	80.9 &	80.7 &	0.868 &	0.25 \\
A4 (9-11) & 81.6 &	82.0 &	81.4 &	0.878 &	0.275 \\
A5 (12-17) & 77.0 &	77.0 &	77.0 & 	0.851 &	0.245 \\
A6 (7-8)   & 69.2 &	69.8 &	69.0 &	0.782 &	0.15 \\ \hline
Female   & 78.3 &	76.8 &	79.2 &	0.855 &	0.24 \\
Male & 76.8 &	76.7 &	76.8 &	0.856 &	0.24 \\ \hline
\end{tabular}
\caption{Results for the different data sets.}
\label{tab:results}
\end{center}
\end{table}

\begin{figure}[ht!]
	\centering
	\includegraphics[width=10cm]{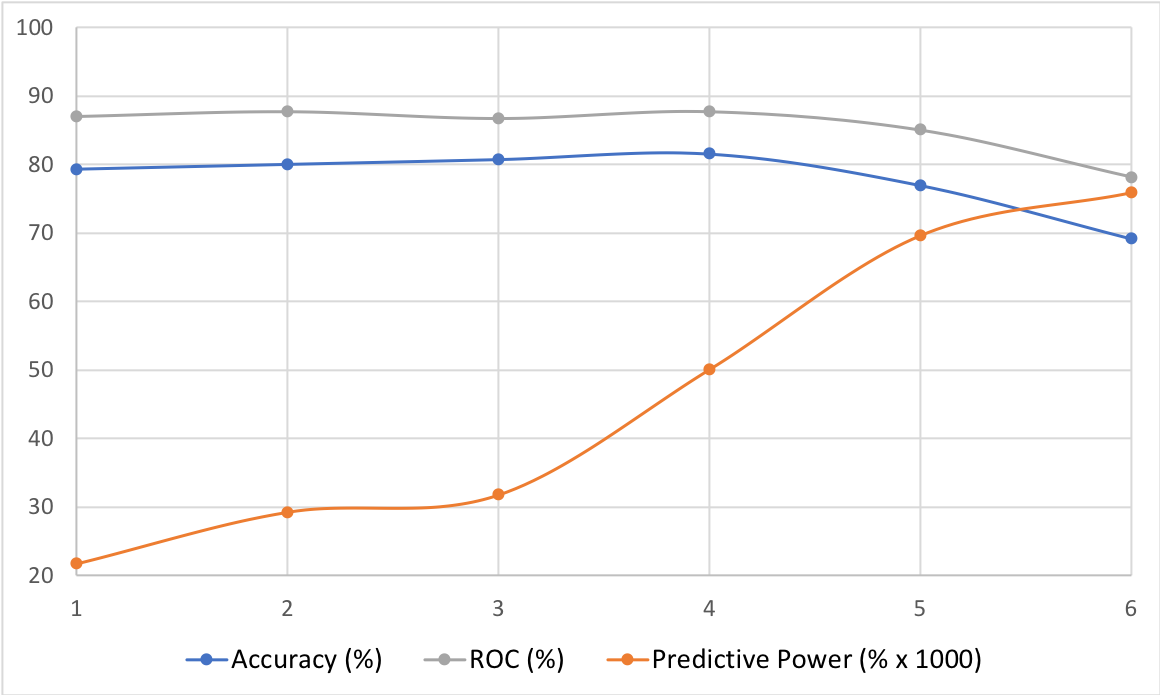}
     \caption{Accuracy, ROC, and predictive power for the different data sets.}
	\label{results}
\end{figure}

In Table~\ref{tab:results} we give the combined accuracy for both classes, the recall and precision for the dyslexia class, the ROC and the threshold used for the Random Forest to obtain these results. In Figure~\ref{results} we show the accuracy, the ROC ({\it i.e.} the Receiver Operating Characteristic), and the predictive power (percentage of accuracy per 1000 people), which shows that about 1,500 participants are enough to reach the accuracy plateau. That implies that doubling the size of A6, would significantly improve the accuracy of the model for younger children.

We also trained classifiers for only the female and male participants finding that with a 5-fold evaluation (validation sets of 20\%), the results were very similar. They are also shown in Table~\ref{tab:results}. 

\subsection*{Deploying the model}

In practice what is important is not the accuracy but the recall (or sensitivity) of the dyslexia class (or the complement of this which is the rate of false positives, that is, the fraction of people without dyslexia that are predicted to have dyslexia) and the rate of false negatives (that is, the fraction of people with dyslexia predicted as without having dyslexia, which is the complement of the specificity), as the impact of each type of error is different. Indeed, missing a child that may have dyslexia is much worse than sending a child without dyslexia to a specialist. 

\begin{figure}[ht!]
	\centering
	\includegraphics[width=10cm]{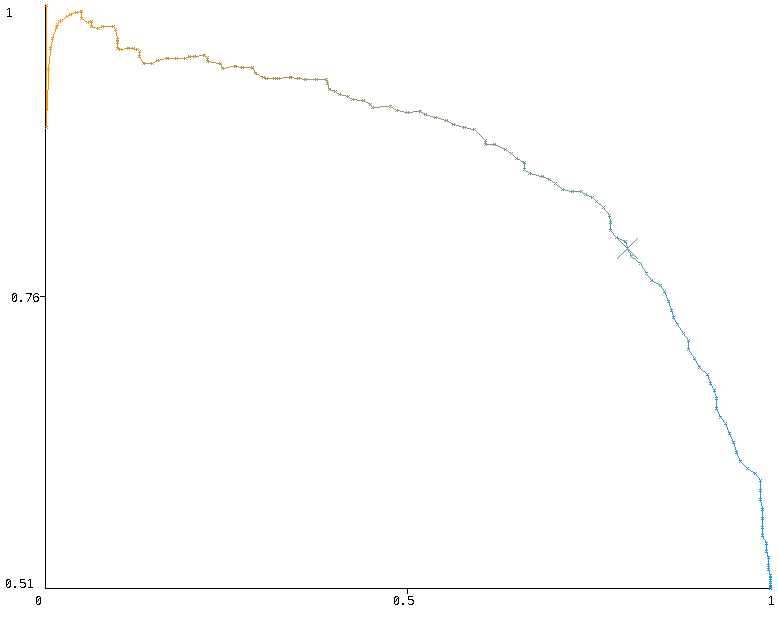}
	\vspace*{0.2mm}
     \caption{Precision and recall curve for the dyslexia class, varying the model threshold.}
	\label{fig:prgraph}
\end{figure}

If the whole population would take the test and the rate of both errors is similar, false negatives would be 9 times more frequent than false positives if we assume a 10\% prevalence. However, in practice, only people that have learning problems takes the test, and we estimate that they are about 20\% of the population \cite{IDAFAQ}. In this case, not only the rate but also the number of false negatives and positives would be similar. Hence, we decided to set the threshold for the model when both types of errors have the rate as similar as possible (0.24 for the main predictive model). 
Our estimation that 20\% of the people who take the test having dyslexia has been proven realistic, as 51\% of the people taking the test are predicted to have risk of dyslexia, which implies a prevalence of 10.2\%. In Table~\ref{tab:confusionmatrix}, we show the precision and recall results per class while in Figure~\ref{fig:prgraph}, we show the precision-recall graph for the dyslexia class, where the point for the threshold of 0.24 is shown with an X.

\begin{table}[ht!]
\begin{center}
\begin{tabular}{|l|c|c|} \hline
Class  &	  Dyslexia (\%)  & No Dyslexia (\%)   \\ \hline
Precision   &  79.7  & 79.1  \\
Recall    &    80.4  & 78.4 \\ \hline
\end{tabular}
\caption{Model precision and recall per class for a threshold of 0.24.}
\label{tab:confusionmatrix}
\end{center}
\end{table}

\subsection*{Over-fitting}

\begin{figure}[ht!]
	\centering
	\includegraphics[width=10cm]{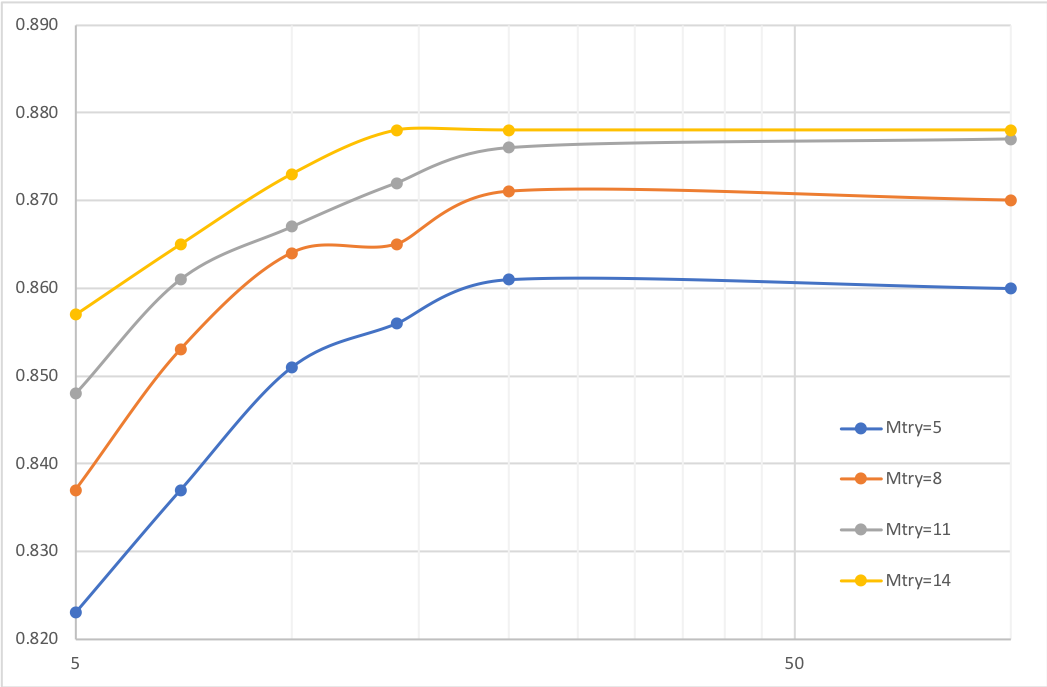}
	\vspace*{0.2mm}
     \caption{ROC in function of two Random Forest parameters for the main model.}
	\label{fig:tuning}
\end{figure}

In spite that, as mentioned before, we designed our models to avoid over-fitting, we did an extra experiment tuning two parameters of the Random Forest: the depth of the tree and mtry, {\it i.e.}, the number of features randomly sampled at every stage. Figure~\ref{fig:tuning} plots the ROC depending on the depth of the tree from 5 to 100 levels and mtry for four values between 5 and 14 where 8 is the default value in Weka and 14 is the default value in R (this value depends on the number of features in both cases). As we can see at depth 20 the result does not improve any longer and even using mtry 14 only improves the ROC marginally. In fact, a model using depth 20 and mtry 14 only obtains a ROC of 0.875 with an accuracy of 79.8\% and sensitivity as well as precision of 79.8\% for a threshold of 0.245. This reaffirms that there is no over-fitting.

\section*{Discussion}
In this section, we explore the impact of the different features and discuss the model’s limitations.

\subsection*{Feature analysis}
To analyze which were the best features in our models, we used standard information gain in decision trees. For example, for the main model (A1), the two most important features were gender and the performance in Spanish classes, which makes sense given that dyslexia is more salient in males and people with dyslexia fail at school. The next 44 best features were related to some question in the test. However, as questions are atomic, we need to aggregate all the features per question. Table~\ref{questions} give the normalized importance (100 the top one) of them, where we also aggregated all the demographic features. 
Notice that all questions discriminate and just using the top features does not give good results. For example, using the top 7 questions plus the demographic variables, the accuracy obtained is just 70.9\%. In Table~\ref{features}, we aggregate features by type, where we can see that, successes are slightly better than mistakes.

\begin{table}[ht!]
\begin{center}
\begin{tabular}{|l|c|l|c|l|c|} \hline
Question &	\% &	Question &	\% &	Question &	\%  \\ \hline\hline
Q1   & 100.0   &    Q10	  & 75.8   &  Demog.  &	67.2  \\
Q3   &	100.0  &	Q13	  & 75.8   &	Q23   &	64.3  \\
Q2   &	98.0   &	Q17   &	75.8   &	Q26   &	64.3  \\
Q4   &	79.2   &    Q21   & 75.4   &    Q24   & 61.5  \\
Q5   &	89.3   &	Q16   &	73.4   &	Q27   &	61.1  \\
Q6   &	87.3   &	Q19   &	71.7   &	Q30   &	60.7  \\
Q7   &	85.7   &	Q18   &	71.3   &	Q25   &	60.2  \\
Q8   &	85.7   &	Q12   &	70.1   &	Q31   &	60.2  \\
Q9   &	84.8   &	Q15   &	69.3   &	Q32   &	60.2  \\
Q14   &	79.9   &	Q22   &	69.3   &	Q29   &	59.8  \\
Q11   &	79.5   &	Q20   &	68.4   &	Q28   &	59.4  \\ \hline
\end{tabular}
\caption{Relative question importance based on feature analysis.}
\label{questions}
\end{center}
\end{table}

\begin{table}[ht!]
\begin{center}
\begin{tabular}{|l|c|l|c|} \hline
Type   &	  \%  & Type   &	  \%  \\ \hline\hline
Hits   &	 100.0 & Accuracy   &  97.3  \\
Score   &	  99.6 &  Miss rate   & 90.1 \\
Misses   &	  99.3 & Demography   &	  16.5  \\
Clicks    &  97.4  & &  \\ \hline
\end{tabular}
\caption{Relative importance by feature type aggregation.}
\label{features}
\end{center}
\end{table}

The most informative features were the set of features coming from the first set of nine questions (Q1-Q9). This is coherent with the design of the test since we placed the questions which were most linguistically motivated at the beginning of the test. Questions 1 to 9 target the basic prerequisites for reading acquisition, such as {\it Phonological Awareness}, {\it Alphabetic Awareness}, {\it Syllabic Awareness} as well as {\it Auditory and Visual discrimination and categorization}. These features come from exercises where the participant was required to  associate a letter name, a letter sound or a syllable with its corresponding spelling. This is consistent with previous literature on dyslexia that specifically focus on the deficit on the phonological component in dyslexia \cite{Lyon2003,Vellutino2004}.

Regarding the aggregated performance measures (Table~\ref{features}), all are highly predictive since all of them address manifestations of the  participant's performance. The feature {\it Hits} is the most predictive one, most likely because the people with dyslexia perform worse.

\subsection*{Customized Test with a New Population}

After these first results, we analyzed how appropriate were the level of the questions for the participants age and how difficult was the test's user interface. As a result, we adapted the test for different age ranges since some questions were too difficult for the younger ages. Hence, in the revised test there are 3 ranges of ages (i) from 7 to 8 years old (327 children) with 19 questions (118 features coming from Q1-Q12, Q14-Q17, Q22-Q23 and Q30); (ii) from 9 to 11 years old (567 children) with 27 questions (166 features coming from Q1-Q20, Q22-Q24, Q26-Q28 and Q30); and (iii) from 12 to 17 years old (498 children) with 31 questions (190 features coming from Q1-Q28 and Q30-Q32). We removed question 31 because the user interaction needed to solve the exercise (cut the sentences into words) was understood differently among participants, {\it i.e.} some used clicks while others dragged the mouse across the words, leading to inconsistent features.

To test the robustness of the method we collected a new data set using a different device, a tablet. This new data set was composed of 1,395 new participants where 10,6\% had diagnosed dyslexia, applying one of the tests above depending on the age of the participant. 
We used the same procedure and inclusion criteria of the main study, where ten new schools participated in the study and the ages of the participants also ranged from 7 to 17.
The participants without dyslexia consisted of 1,247 people (M = 10.75 years old, SD = 2.46) being 50.04\% female and 49.96\% male. The group of participants with dyslexia was composed of 148 people (M = 9.61 years old, SD = 2.11), where 51.4\% were female and 48.6\% male.

Table~\ref{tablet} shows the data set characteristics and the results for the three different age groups. As we can see, for children 9 years old or older, we obtain almost 75\% sensitivity in spite of using less participants and features.

\begin{table}[ht!]
\begin{center}
\begin{tabular}{|l|c|c|c|c|c|c|c|} \hline
Data set & Accur. & Recall & Precis. & Recall & Precis. & ROC & Thresh. \\ 
(age range) & (\%) & (Dys, \%) & (Dys,\%)& (Non, \%) & (Non,\%)& & \\ \hline\hline 
N1 (12-17)  & 74.9	& 75.0 & 75.0 & 75.0 & 75.0 & 0.805 &	0.15 \\
N2 (9-11)  & 72.6 & 72.5 & 72.8 & 72.8 & 72.5 & 0.790 & 0.225 \\
N3 (7-8)  & 60.8 &	61.1 & 60.6	& 60.4 & 60.7 & 0.672 & 0.29 \\ 
\hline
\end{tabular}
\caption{Results for the tablet test.}
\label{tablet}
\end{center}
\end{table}

\subsection*{Limitations}
Our machine learning model trained from human-computer interaction data is able to classify people as having dyslexia or not with high sensitivity, and using this type of data to screen dyslexia is novel. However, it indirectly considers measures that have previously used in traditional diagnoses. Indeed, paper based tests use reading and writing performance measures such as reading speed, spelling errors, and text comprehension \cite{TALE1984,PROLECR,DSTJ}, and the measures gathered with our online test indirectly measures such user's performance when the participant is exposed to the linguistic questions.

Nevertheless, the results of this online test should be taken as screening only and cannot serve as a diagnosis due to at least three reasons. First, our online test does not take into consideration other factors such as the intelligence quotient of the participant. In professional practice, intelligence tests such as WISC \cite{WISC}, are normally taken in order to diagnose dyslexia and exclude other possible causes of the phonological skills deficiencies. 

Second, our test does not discriminate other conditions. It is increasingly recognized that dyslexia co-occurs with other disorders \cite{snowling2013early}. For instance, dyslexia is often co-morbid with dyscalculia \cite{gross1996developmental} and attention deficit hyperactivity disorder (ADHD) \cite{Pauc2005}. Notably, 40\% of the people with dyslexia have dyscalculia \cite{wilson2015dyscalculia}, and from 18 to 42\% of the population with dyslexia also have attention deficit hyperactivity disorder (ADHD) \cite{Pauc2005}. Also, there are other language disorders, such as specific language impairment (SLI), that require professional assessment. 
These comorbidities make professional diagnoses a more challenging task, and, in practice, sometimes dyslexia is misdiagnosed by ADHD and {\it vice versa} \cite{Dakin2005,Tridas2007}. In our approach we took as ground truth the current dyslexia diagnosis accessed by a professional, however, that ground truth could vary depending on the professional assessment. Furthermore, there can be other factors that can play a role such as fatigue and concentration. 

Finally, our test cannot report different degrees of dyslexia and does not consider the personal history of the user which can also play a role on dyslexia diagnosis.

\section*{Conclusions}
The approach presented in this article shows that dyslexia can be screened in a language with shallow orthography, such as Spanish, using machine learning in combination with measures derived from a gamified online test. However, the results of this approach should be taken as a screening test in practice, never as a dyslexia diagnosis, since there are other factors such as intelligence quotient and dyslexia comorbidities that needs professional oversight. 

This approach of screening dyslexia is easy to take on the Web, since it does not require special equipment. So far, our test has been deployed as an open access on-line tool used already more than 200,000 times in Spanish speaking countries. Since estimations of dyslexia are much higher than the actual diagnosed population, we believe this method has potential to make a significant social impact. Similar methods could lead to earlier detection of dyslexia and prevent children from being diagnosed with dyslexia only after they fail in school.

Nevertheless, we need to carry out further experiments with the new tests for tablets as well as to collect larger data sets for building more accurate models.

\section*{Acknowledgments}
\paragraph{Funding:} Financial support was provided by a grant from the US Department of Education NIDRR (grant number H133A130057, J.B., https://www.ed.gov/); and a grant from the National Science Foundation (grant number IIS-1618784, J.B. and L.R., https://www.nsf.gov/).

We thank the volunteers that participated in this study,  the voice actress Nikki Garc\'ia, and the speech therapists who reviewed the test, 
Alicia Bailey Garrido, 
Daniel Cubilla Bonnetier,
Nancy Cushen-White,
Ruth Rozensztejn, and 
Daniela S\'anchez Alarc\'on.

As well as the professionals from the educational centers who helped reviewing some cases of the ground-truth from the data set:
\'Angeles \'Alvarez-Cedr\'on
Angela Biola Quintana P\'erez,
Mireia Centeno,
Patricia Clemente,
Pilar Del Valle Sanz,
Esther Gamiz,
Paloma Garc\'ia Rodr\'iguez,
Jos\'e Manuel Gonz\'alez Sanz,
Cristina Mart\'in,
Miguel \'Angel Matute, and
Ana Olivares

We thank the specialized centres for providing participants with diagnosed dyslexia: 
Academia Eklekticos,
ADAH SLP Vigo,
Apr\`en$+$, 
Atenea Psicosalud \& Psicoeducativo,
CAMINS Logop\`edia, Psicologia i Dificultats d'Aprenentatge,
Centre Elisenda Curri\`a,
Centre Espais,
Centre Neureduca,
Claudia Squella, 
CREIX Centre d'Assessorament Psicopedag\`ogic Barcelona,
CREIX – Centro de Desarrollo Infantil Mallorca,
Did\`actica-Rub\'i,
Educasapiens,
Engracia Rodr\'iguez-L\'opez Domingo,
Gabinete Psicolog\'ia Marta Pellejero Escobedo,
Isabel Barros,
Logop\`edics Lleida,
Novacadèmia from Barcelona, Sant Feliu de Codines and Caldes de Montbui,
Tangram Barcelona,
Uditta,
UTAE (Unitat de Trastorn de l'Aprenentatge Escolar), 
and
Valley Speech Language and Learning Center Texas.

We thank the following non-profit organizations for providing participants and spreading the call for participation:
ADA Dislexia Arag\'on,
Adixmur,
Asociaci\'on ACNIDA,
Asociaci\'on Catalana de la Dislexia,
Associació de Disl\`exia Lleida,
COPOE (Confederaci\'on de Organizaciones de Psicopedagog\'ia y Orientaci\'on de Espa\~na),
COPOE (Orientaci\'on y Educaci\'on Madrid),
Disfam,
Disfam Argentina,
Dislexia \& Dispraxia Argentina, 
Fundaci\'o Mirades Educatives.
Fundaci\'on Educere,
Fundaci\'on Marillac,
Fundaci\'on Vals\'e,
and
Madrid con la Dislexia.

We are also very grateful to the schools and the universities that participated in the main study: 
CEIP Bisbe Climent,
CEIP Foro Romano,
CEIP Juan XXIII,
CEIP Los \'Angeles,
CEIP Maestro Juan de \'Avila,
CEIP Ntra. Sra. de la Salud,
CEIP Nuestra Se\~nora de los \'Angeles,
CEIP San Jos\'e de Calasanz Fraga,
CEIP San Jos\'e de Calasanz Getafe,
CEPA Ignacio Zuloaga Helduen Heziketa Iraunkorra,
CES Vega Media,
Colegio Adventista Rigel,
Colegio Alborada,
Colegio Am\'erico Vespucio,
Colegio Areteia,
Colegio Concertado Biling\"ue Divina Providencia,
Colegio de Fomento Las Tablas-Valverde,
Colegio de las Hermanas de la Caridad de Santa Ana,
Colegio Gimnasio los Pinares,
Colegio Hijas de San Jos\'e,
Colegio La Milagrosa,
Colegio Madre Paulina de Chiguayante,
Colegio Mar\'ia Auxiliadora de Alicante,
Colegio Mar\'ia Auxiliadora de Sep\'ulveda,
Colegio Mar\'ia Auxiliadora de Sueca,
Colegio Mar\'ia Auxiliadora de Terrasa,
Colegio Mar\'ia Auxiliadora de Torrent,
Colegio Mar\'ia Auxiliadora de Valencia,
Colegio Mar\'ia Auxiliadora de Zaragoza,
Colegio Mar\'ia Inmaculada de Concepci\'on,
Colegio Mar\'ia Moliner,
Colegio Matilde Huici Navas,
Colegio Miguel Servet,
Colegio Nuestra Señora de la Soledad,
Colegio Obispo Perell\'o,
Colegio Rural Agrupado Tres Riberas,
Colegio San Gabriel,
Colegio Santa Ana Fraga,
Colegio Santa Ana Zaragoza,
Colegio Santa Dorotea,
Colegio Santa Mar\'ia del Pilar Marianistas de Zaragoza,
Colegio Virgen de la Pe\~na,
CPI Castroverde,
Escola 4 Vents,
Escola Comptes de Torregrossa,
Escola Les Cometes,
Escola Mare de D\`eu del Priorat,
Escola Pepa Colomer,
Escola Sol Ixent,
GSD Guadarrama,
IES Azuer,
IES Bajo Cinca,
IES Ben Gabirol,
IES Corona de Arag\'on en Zaragoza,
IES do Cami\~no,
IES Leonardo de Chabacier,
IES Puerta del And\'evalo,
IES Ram\'on J. Sender,
IES Rey Fernando VI,
IES de Bocairent,
University of Valencia, 
University Don Bosco, 
and 
University San Jorge.

Finally, we thank the schools who participated in the second user study:
Centro Infanta Leonor,
Colegio Sagrado Coraz\'on,
Colegio San Antonio,
Colegio San Patricio,
Colegio San Prudencio,
Colegio Santo Domingo,
Colegio Urkique,
Colegio Vizcaya,
Escolapios de Getafe, and
Escuelas Bosque.


\section*{Supporting information}
\paragraph{Data Availability:} The data sets reported in this work are archived and freely accessible at
\url{https://datadryad.org/stash/share/PGE-YuBa0be3tq0I1wedR-7hgko-qTAan3XRjL-0DOc}

The full gamified questions of the test are available at \url{https://www.playdyslexia.org/}.
The test has been deployed as an open access on-line tool available at \url{https://dytectivetest.org}.

\end{document}